\documentclass[
amsmath,amssymb,
aps,
prd, 
twocolumn, 
amsmath,  
nofootinbib,
superscriptaddress,
showkeys
]{revtex4-2}
\usepackage[utf8]{inputenc}
\usepackage{graphicx}
\usepackage[T1]{fontenc}
\usepackage{float}
\usepackage{color}
\usepackage{hyperref}
\makeatletter\def\Hy@Warning#1{}\makeatother
\usepackage{subfiles} 
\usepackage[normalem]{ulem}
\usepackage{bm}
\usepackage{float}
\usepackage{mathtools}
\DeclarePairedDelimiter\abs{\lvert}{\rvert}
\hypersetup{colorlinks=true, citecolor=blue}

\newcommand{\pr}[1]{\ensuremath{\left[#1\right]}}

\newcommand{\chav}[1]{\ensuremath{\left\{#1\right\}}}

\begin{document}

\title{
Detecting the third family of compact stars with normalizing flows
}

\author{Valéria Carvalho}
\email{val.mar.dinis@uc.pt}
\affiliation{CFisUC, 
	Department of Physics, University of Coimbra, P-3004 - 516  Coimbra, Portugal}

\author{Márcio Ferreira}
\email{marcio.ferreira@uc.pt}
\affiliation{CFisUC, 
	Department of Physics, University of Coimbra, P-3004 - 516  Coimbra, Portugal}
	
\author{Constança Providência}
\email{cp@uc.pt}
\affiliation{CFisUC, 
	Department of Physics, University of Coimbra, P-3004 - 516  Coimbra, Portugal}

\author{Micha{\l} Bejger}
\email{bejger@camk.edu.pl}
\affiliation{INFN Sezione di Ferrara, Via Saragat 1, 44122 Ferrara, Italy}
\affiliation{Nicolaus Copernicus Astronomical Center, Polish Academy of Sciences, Bartycka 18, 00-716, Warsaw, Poland}

\date{\today}

\begin{abstract}
We explore the anomaly detection framework based on Normalizing Flows (NF) models introduced in \cite{PhysRevC.106.065802} to detect the presence of a large (destabilising) dense matter phase transition in neutron star (NS) observations of masses and radii, and relate the feasibility of detection with parameters of the underlying mass-radius sequence, which is a functional of the dense matter equation of state. Once trained on simulated data featuring continuous $M(R)$ solutions (i.e., no phase transitions), NF is used to determine the likelihood of a first-order phase transition in a given set of $M(R)$ observations featuring a discontinuity, i.e., perform the anomaly detection. Different mock test sets, featuring two branch solutions in the $M(R)$ diagram, were parameterized by the NS mass at which the phase transition occurs, $M_c$, and the radius difference between the heaviest hadronic star and lightest hybrid star, $\Delta R$. We analyze the impact of these parameters on the NF performance in detecting the presence of a first-order phase transition. Among the results, we report that given a set of 15 stars with radius uncertainty of $0.2$ km, a detection of a two-branch solution is possible with 95\% accuracy if $\Delta R > 0.4$ km.  
\end{abstract}

\maketitle


\section{\label{sec:introduction} Introduction}
Discovering the properties of the dense nuclear matter realized inside neutron stars (NS) is yet a fundamental unsolved question in nuclear physics. The present constraints on the high density nuclear matter equation of state (EoS) are originated from  NS observations. The existence of hybrid stars, where a phase transition from nuclear matter to deconfined quark matter in the star's core occurs, is yet to be validated. The nature of the hypothetical phase transition depends on the surface tension between both phases, which could be either a sharp (Maxwell construction) for large values, or a smooth mixed phase transition for low values \cite{Alford:2013aca,Alford:2015gna}. Assuming a large surface tension value, the transition between nuclear and quark phases is thus realized via a first order phase transition. For some hybrid stars models, depending on the energy density discontinuity between both phases, a disconnected branch may appear in the $M(R)$ diagram \cite{Schertler:2000xq,Alford:2013aca,Benic:2014jia,Alford:2015gna,Alvarez-Castillo:2018pve,Blaschke:2020qqj}. The value of the energy density discontinuity dictates the stability of the hybrid star: for low values, the quark core is able to counteract the gravitational pressure from the nuclear mantle, otherwise, a too strong energy density discontinuity, destabilizes the hybrid star. 
By analyzing the qualitative features induced in the mass-radius diagram, 
this work aims to analyze the detectability of such first-order phase transitions, specifically when the stable branch of hybrid stars is disconnected from the hadronic branch is realized, and assessment of criteria - number of observations, observational errors, parameters of the phase transition -- at which the detectability statement is credible.\\

Generative machine learning models are promising tools in tackling several challenges in high energy physics (see \cite{Zhou_2024}), e.g., in anomaly detection tasks in particle physics problems \cite{Belis:2023mqs}.
Among the different generative models, normalizing flows (NF) models \cite{rezende2016variational,Kobyzev_2021} provide an analytic description of the underlying data distribution, simultaneously enabling efficient sampling and probability density estimation. 
In astrophysics, NF have been employed for 
rapid gravitational-wave parameter estimation parameters from gravitational-wave data \cite{PhysRevD.102.104057,PhysRevD.107.084046}, 
 showing the remarkable versatility of this probabilistic generative model. 
The recent study \cite{Brandes:2024vhw} combines NF with Hamiltonian Monte Carlo methods to infer the complete posterior distribution of the EoS and nuisance parameters directly from telescope observations.
Additionally, a novel framework for detecting possible phase transitions based on NSs mass-radius  
observations was introduced in \cite{PhysRevC.106.065802}. The authors employed a flow-based model for anomaly detection, specifically testing NF models on multiple samples of six EoS exhibiting different phase transitions. The evaluation involved the analyses of the latent space to determine if the model successfully detected anomalies. \\

The present work extends the NF methodology introduced in \cite{PhysRevC.106.065802}, by applying it to generic $M(R)$ diagrams instead of specific hybrid EoS. The generic $M(R)$ diagrams are constructed to simulate phase transitions at a specific NS mass and with a given radius difference between the heaviest hadronic star and lightest hybrid star. Furthermore, we use directly the NF ability to perform probability density estimation on input vectors, determining whether anomalies are present in the input vectors. \\
  
The structure of this paper unfolds as follows: In Sec.~\ref{sec:nf}, we provide an introduction to NF. Section \ref{sec:dataset} introduces the dataset used, the underlying theory, and the process of implementing mock observations. 
Section \ref{sec:results} scrutinizes the obtained results, elucidating the metrics and datasets employed. Finally, Sect.~\ref{sec:conclusions} summarizes our findings and concludes the study.

\section{\label{sec:nf} Normalizing flows}
\begin{figure*}[!htb]
    \centering
    \includegraphics[width=0.9\linewidth]{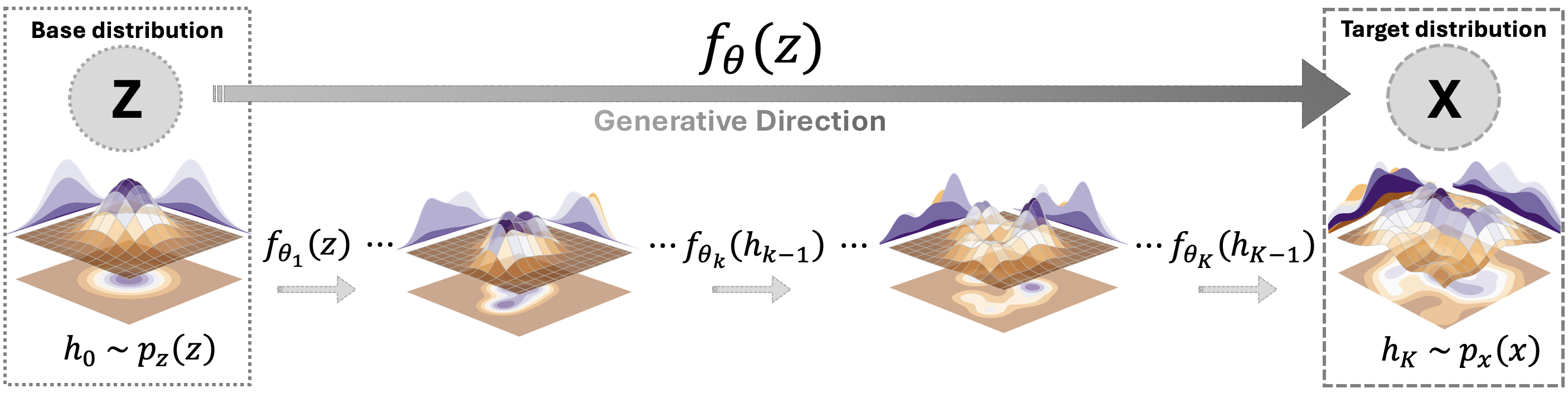}
    \caption{Schematic representation of a flow-based model illustrating the application of a transformation $\mathbf{f}_\theta(\mathbf{z})$ across $K$  coupling layers, denoted as $\mathbf{f}_{\theta_K}\circ\cdots \circ \mathbf{f}_{\theta_2} \circ \mathbf{f}_{\theta_1}(h_0)$. This process operates on samples from a base distribution $p_z(\mathbf{z})$, yielding samples of the target distribution $p_x(\mathbf{x})$ in the generative direction. The opposite direction is known as the normalizing direction $\mathbf{f}^{-1}_\theta(\mathbf{x})$. The dataset elements $x_i\sim X$ are composed of $N$ mass-radius observations, 
$x_i=\pr{(M_1,R_1),(M_2,R_2),...,(M_N,R_N)}$, see Sec.~\ref{sec:dataset} for details.}
    \label{fig:NF}
\end{figure*}

Normalizing flows (NF), extensively reviewed in \cite{papamakarios2021normalizing}, constitute a probabilistic generative model based on invertible transformations, which has been explored in physics, e.g., \cite{PhysRevC.106.065802,PhysRevD.102.104057,PhysRevD.107.084046}. 
Their goal is to model the intricate data probability distribution $p_x(\mathbf{x})$
 by transforming random variables $\mathbf{z}$ from a simple base distribution $p_z(\mathbf{z})$—typically a multivariate standard normal distribution—through a nonlinear yet invertible and differentiable bijective transformation $\mathbf{f}_\theta: \mathbb{R}^N \to  \mathbb{R}^N $. These transformations are defined by neural networks (see \cite{Zhou_2024} for a comprehensive review with applications in physics) and parametrized by $\theta$. The transformation can be expressed as
 \begin{equation}
	\mathbf{x}= \mathbf{f}_\theta(\mathbf{z}), \quad \text{where} \quad \mathbf{z} \sim p_z(\mathbf{z}) \; .
 \end{equation}
Applying the change of variables formula from probability theory allows us to compute the probability distribution
\begin{align}
	p_x(\mathbf{x})=  p_z(\mathbf{z}) \abs*{\text{det} \frac{ \partial \mathbf{f}_\theta(\mathbf{z})}{\partial \mathbf{z}} }^{-1}=p_z(\mathbf{f}^{-1}_\theta(\mathbf{x})) \abs*{\text{det} \frac{\partial \mathbf{f}^{-1}_\theta(\mathbf{x})}{\partial \mathbf{x}} }\; .
\end{align}
The second term represents the absolute value of the Jacobian determinant, where $\partial \mathbf{f}^{-1}_\theta(\mathbf{x})/\partial \mathbf{x}$ is an $N \times N$ matrix. This determinant captures the change in volume resulting
from the transformation applied to the probability space. The transformation itself serves to morph the base distribution into the target distribution.\\\\
The mapping $\mathbf{f}_\theta$ can be expressed as a composition of invertible functions: $\mathbf{f}_{ \mathbf{\theta}}=\mathbf{f}_{\theta_K}\circ\cdots \circ \mathbf{f}_{\theta_2}  \circ \mathbf{f}_{\theta_1}$.
Consequently, the target variable is obtained through the composition:

\begin{equation}
	\mathbf{x}= \mathbf{f}_\theta(\mathbf{z})= \mathbf{f}_{\theta_K}\circ\cdots \circ \mathbf{f}_{\theta_2}  \circ \mathbf{f}_{\theta_1}(\mathbf{z})\; .
\end{equation}
The relation between $\mathbf{x}$ and $\mathbf{z}$ is then 
$\mathbf{z} \xrightarrow{\mathbf{f}_{\theta_1}}h_{1} \xrightarrow{\mathbf{f}_{\theta_2}} \cdots \xrightarrow{\mathbf{f}_{\theta_{K-1}}} h_{K-1}  \xrightarrow{\mathbf{f}_{\theta_{K}}}\mathbf{x} $,
where $h_0=\mathbf{z} $ and $h_K=\mathbf{x}$. The schematic representation of a flow-based model is in Fig.~\ref{fig:NF}.
The logarithm of the probability distribution can then be defined as:

\begin{align}
	\text{log } p_x(\mathbf{x})   = \text{log } p_z(\mathbf{f}^{-1}_{\theta_k}( h_{k-1})) + \sum_{k=1}^K \text{log }  \abs*{\text{det} \frac{\partial \mathbf{f}^{-1}_{\theta_k}( h_{k-1})}{\partial h_{k-1}} } \; .
\end{align}
The term $\text{log } p_z(\mathbf{f}^{-1}_{\theta_k}( h_{k-1})) $ captures the logarithm of the base distribution's probability, while the summation over the
$K$ terms represents the cumulative contribution of Jacobian determinants from each invertible function in the composition. 
The overall likelihood of the data under the NF model is tractable and leads to
the training objective (loss function) being minimizing:

\begin{align}\label{eq:NLL}
	\mathcal{L}(\mathcal{D})=- \frac{1}{\abs*{\mathcal{D}}} \sum_{\mathbf{x}\epsilon \mathcal{D}}  \text{log } p_x(\mathbf{x}) \; ,
\end{align}
which is the negative log-likelihood (NLL) for the training dataset $\mathcal{D}$.\\

While various implementations of NF exist, we select the coupling neural spline flow \cite{durkan2019neural} for the present work. 
In the context of coupling transforms, the input variable is divided into two parts, [$x_{1:k-1}$, $x_{k:K}$], where $\mathbf{f}_\theta$ is applied to the second part, resulting in an output of [$x_{1:k-1}$,$\mathbf{f}_\theta (x_{k-1:K})$]. Values are shuffled between each transformation using a 
   permutation layer. For the actual transformation, we utilized a rational-quadratic spline function.
For the actual implementation, we have used the \texttt{PyTorch} \cite{NEURIPS2019_9015} supplemented with the \texttt{nflows} library \cite{conor_durkan_2020_4296287}.

\section{\label{sec:dataset} Dataset}

The possible existence of a first-order phase transition between nuclear and quark matter in NS remains an open question. 
Due to the uncertainty on the value of the surface tension at the interface, the phase transition between two pure phases, say nucleonic and quark could be either sharp (so-called Maxwell construction, featuring a large discontinuity in density profile) or a smooth one (via the ``mixed phase'', so-called Gibbs construction) \cite{Yasutake:2014oxa}.
Assuming a high surface tension scenario, and thus a sharp phase transition, the appearance of a disconnected branch of hybrid stars depends on the energy density discontinuity value $\Delta \epsilon$ at the transition \cite{Alford:2013aca}.\\

\begin{figure}[!htb]
    \centering    \includegraphics[width=1.0\linewidth]{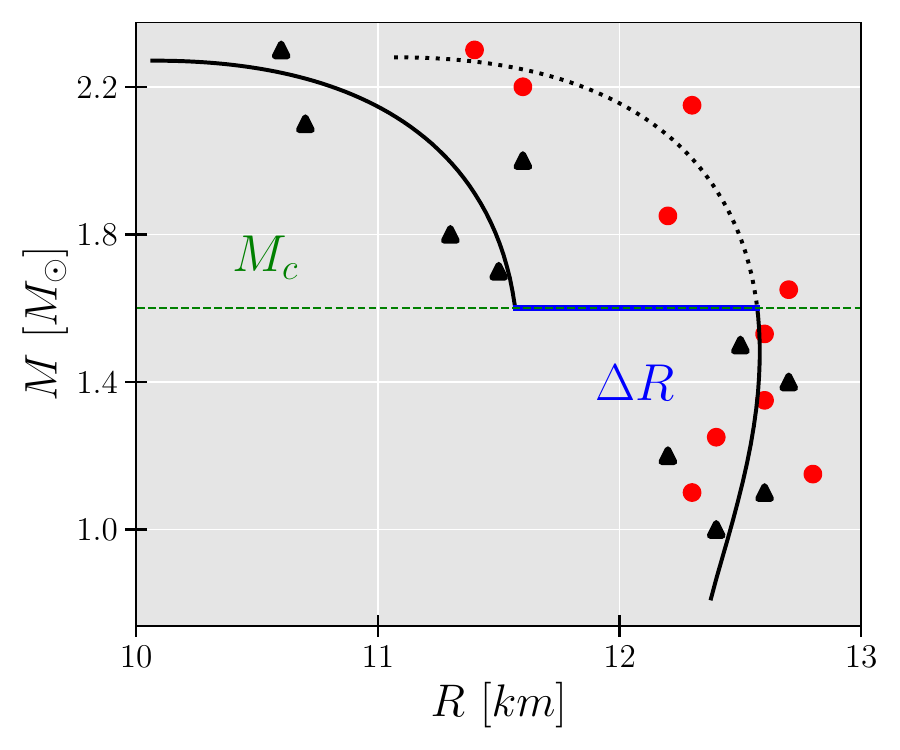}
    \caption{To simulate the existence of two branches in a given TOV solution, we have defined a critical mass $M_{\text{c}}$, and the upper part solution of a hadronic EoS was shifted by $\Delta R$. To illustrate the sampling procedure described in Sec.~\ref{generation}, we display a sample from the purely hadronic EoS (red dots) and from the hybrid $M(R)$ solution (black triangles).  }
    \label{fig:diagram}
\end{figure}

The present work investigates the hypothesis of the existence of hybrid star branches, and aims to analyze their detectability from a set of NS observations. Since we are only interested in accessing the degree of detectability of such branches regardless of the details of both nuclear and quark matter phases, we constructed a mock dataset that simulates a continuity of solutions. We used, as a starting point, a set of hadronic EoS and emulate the appearance of a second branch solution in the $M(R)$ diagram as illustrated in Fig.~\ref{fig:diagram}. We selected a NS mass $M_{\text{c}}$ and shifted the upper section by $\Delta R$, which simulates the difference in radius between the hadronic star and a hybrid star (with quark matter core). This construction allows one to analyze the impact of $(M_{\text{c}}, \Delta R)$ in the likelihood of observing such scenarios given a set of $M(R)$ observations. Notice that this hybrid star mass-radius curve construction may be considered conservative because in a realistic description of a quark branch it is expected that: i) the second stable branch may start at $M<M_c$ as in \cite{Benic:2014jia}; ii) the slope of the mass-radius of the second branch will probably be quite different from the one of the hadronic mass-radius behavior. We will test our model with a mass-radius curve obtained from microscopic models. 

The key insight of using NF in the present task is the following: once NF models are trained with samples from $M(R)$ diagrams originated by hadronic EoS (red dots in Fig.~\ref{fig:diagram}), we are capable of estimating the likelihood of any other $M(R)$ diagrams, such as the sample from a two-branch solution (black triangles in Fig.~\ref{fig:diagram}) -- a well trained model attributes low likelihood to any samples that deviate considerably from the statistical properties of the training set.   

The base set of hadronic EoS consists 25 287 nuclear models based on a relativistic mean field (RMF) description, which we divide in 80$\%$ train and 20$\%$ test, where minimal constraints were imposed \cite{Malik:2023mnx}: several nuclear saturation properties, the existence of a 2$M_\odot$ NS, and a consistent low-density pure neutron matter with N$^3$LO calculations in chiral effective field theory. The specific dataset used in our study corresponds to Set 0 from the article \cite{Malik:2023mnx} and the dataset has been made available in \cite{malik_zenodo} (Fig.~14 in \cite{Malik:2023mnx} shows the full $M(R)$ posterior of this dataset). 

\subsection{Observation mock data\label{generation}}

The generation of the mock datasets used to train the NF models closely follows our previous works \cite{Carvalho:2023ele,Carvalho:2024kgf}. We sample $N$ mass values, $M_n^{0}/M_{\odot}$, from $\mathcal{U}(1,M_{\text{max}})$, where $n=1,...,N$, since no NS for masses below $1\,M_\odot$ are expected from stellar evolution. Then, we determine both $\sigma_{n,M}$ and $\sigma_{n,R}$ from $\mathcal{U}(0, \sigma_M)$ and $\mathcal{U}(0, \sigma_R)$, respectively. The final $\chav{M_n,R_n}$ observations are obtained from $R_n \sim \mathcal{N}\left(R\left(M_n^{0}\right),\sigma_{n,R}^2 \right)$ and $M_n \sim \mathcal{N}\left(M_n^{0},\sigma_{n,M}^2 \right)$. Basically, we are creating a mock observation set with $N$ stars whose observations are Gaussian scattered around the true TOV solution, with specific standard deviations for each star, $\sigma_{n,M}$ and $\sigma_{n,R}$. We generate $n_{\text{s}}=120$ mock observation sets by repeating the described procedure $120$ times for each EoS. 

In total, we have generated three different data sets, the properties of which are detailed in Table \ref{tab:sets}.
While radius uncertainties of $\sim0.1$ km or less are expected from future observations (e.g., Einstein Telescope \cite{Branchesi:2023mws}, Cosmic Explorer’s \cite{Evans:2021gyd}), 
we employed more conservative uncertainty parameters for Set 1: $\sigma_R=0.2$ km and $\sigma_M=0.1M_{\odot}$. Set 2 was derived using the methodology described in our prior article \cite{Carvalho:2024kgf}. 
For each set, we trained two models with $N=5$ and $15$ to investigate how the number of stars affects the model performance in detecting a second branch in the TOV solution.

Let us stress that the NF models were trained in datasets in which only a single hadronic branch exists, i.e., they were trained with $\Delta R=0$. After training, we applied these NF models in test datasets specified by $(M_{\text{c}}, \Delta R)$ in order to access the model performance in detecting the existence of a second brunch. 
Contrarily to $n_s=120$ mock observations used in the training sets, all test sets are composed of $n_s=1$ mock observations to mimic a real-world scenario where we have access to a single mock observation of the "true" EoS.

\begin{table}[!hbt]
\centering
   \caption{Generation parameters for each dataset.}
   \label{tab:sets}
    \begin{tabular}{ccc}
    \hline
    \hline
    Dataset & $\sigma_M \,[M_\odot]$ & $\sigma_R$ [km]   \\ \hline
    \hline
    Set 0  & 0     & 0   \\ \hline
    Set 1  &   0.1  & 0.2    \\ \hline
    Set 2  &  0.136   & 0.626     \\ \hline
    \end{tabular}
\end{table}

\subsection{Training procedure  \label{training}}

The training process was conducted for three distinct datasets (see Table \ref{tab:sets}) and two different input sizes, $N=5$ and $15$, as elaborated in Sec.~\ref{generation}, resulting in the
 training of six distinct models. Throughout the training, the dataset for training was partitioned into $80\%$ for actual training and $20\%$ for validation. 
 A test on data normalization revealed superior results with standardization than with the method employed in the work \cite{Fujimoto:2024cyv} for our case.
Our chosen architecture is a concatenation of blocks featuring an invertible linear transformation using 
LU-decomposition \cite{kingma2018glow} alongside a rational quadratic spline transform \cite{durkan2019neural}
 employing 8 bins. To obtain the parameters of the spline and the respective $\mathbf{f}_\theta$, we utilized a multilayer perceptron. Various configurations were 
 explored, including changes in the number of hidden layers, neurons per layer, and activation functions. Optimal results were achieved with 3 hidden layers, each 
 comprising 15 neurons, and the Softplus function as the activation function. For input instances with five stars ($N=5$), i.e., five $\chav{M_i,R_i}$ pairs, seven coupling transformations were implemented, 
 while for 15 stars ($N=15$), i.e., 15 $\chav{M_i,R_i}$ pairs, we utilized ten.
 The training setup involved a learning rate of 0.001, utilizing the ADAM optimizer \cite{kingma2014adam} with the AMSgrad improvement \cite{reddi2019convergence}.
  The models undergo training for 500 epochs with a batch size of 1024. 

\section{\label{sec:results} Results}

\begin{figure*}[!htb]
    \centering
    \includegraphics[width=1.0\linewidth]{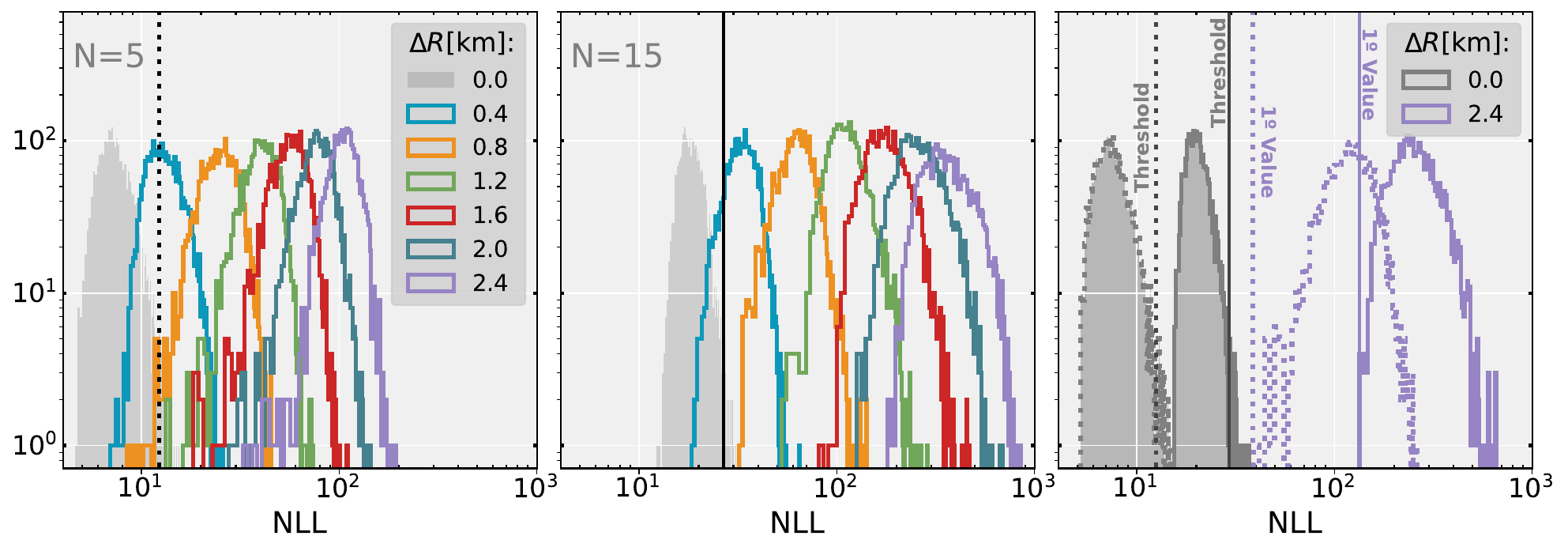}
    \caption{NLL histograms for the $N=5$ (left) and $N=15$ (middle) NF models on the different $\Delta R$ sets (colors) using $M_c=1.5M_{\odot}$. 
    All test sets were generated according to Set 1 properties (see Table \ref{tab:sets}).
    The grey regions represent the NLL of each NF model on the test sets with $\Delta R=0$ (purely hadronic solution) and the vertical black lines correspond to the respective threshold value defined in that region. 
    A comparison between $N=5$ (dotted) and $N=15$ (solid) NF models for $\Delta R=2.4$ km (purple) is also shown (right), where the vertical purple lines mark the first value of the histogram.  }
    \label{fig:hist}
\end{figure*}

After training the NF with NS data with no phase transitions, i.e., with purely hadronic branches, we assess the models performance in several mock datasets featuring varying degrees of phase transitions, as detailed in Sec.~\ref{sec:dataset}. Specifically, we conducted tests with three critical masses ($M_c$) representing phase transitions at 1.2, 1.5, and $1.8M_\odot$, and with radius shifts ($\Delta R$) ranging from 0.0 to 2.4 km, with intervals of 0.2 km (see Fig.~\ref{fig:diagram}). We ensured that each sample from the $M(R)$ diagrams contains at least one star in each branch, i.e., hadronic and hybrid. This procedure was applied across all two-branch solution diagrams within the different $(M_c, \Delta R)$ test sets, containing 2529 instances each.\\

\textit{How can we precisely quantify anomalies in our NF model?} Leveraging the ability of NF to provide direct probability density estimation on target inputs, we anticipate that $\chav{M_n,R_n}$ inputs from two-branch solutions yield lower density values, signalling the presence of an anomaly (phase transition). 
Calculating the NLL (see Eq.~\ref{eq:NLL}) on the test sets, anomalies are anticipated to exhibit substantially high NLL values. To quantify anomalies across different $\Delta R$ values, we approach it as a binary classification problem: the positive class represents the presence of an anomaly, while the absence of an anomaly is the negative class.

We establish the anomaly detection threshold by fixing the False Positive Rate (FPR)\footnote{The FPR is calculated as FPR=${FP}/{(FP+TN)}$, where FP stands for False Positive (the number of actual negatives predicted as positives), and TN corresponds to True Negative (the number of actual negatives predicted as negatives).} at $1\%$. The FPR, a metric that quantifies the model's tendency to incorrectly classify negative instances as positive, thus defines the condition for threshold determination on the test sets for the cases with $\Delta R=0$ (indicating no anomaly).
Since we have three sets (see Table \ref{tab:sets}) and two input vector sizes, $N=5$ and $15$, we have trained a total of 6 NF models, each with its corresponding threshold anomaly.\\

The process of anomaly detecting is illustrated in Fig.~\ref{fig:hist}, where we show the $N=5$ (left panel) and $N=15$ (middle panel) NF models trained on Set 1, using $M_c=1.5 M_\odot$. The grey histograms indicate the NLL of the test sets, and the vertical lines indicate the corresponding threshold values (FPR$=1\%$). Note that both the grey histograms and the thresholds are defined using the test sets of the one-branch solutions (hadronic EoS), i.e., $\Delta R=0$. The other colored histograms represent the NLL of the different $\Delta R$ sets. For instance, the detection rate of any two-branch solution with $\Delta R\geq0.8$ km is 100\% for $N=15$ stars (middle panel).
The right panel represents a comparison between the two NF models on the $\Delta R =2.4$ km (purple) sets. As expected, having access to 15 stars is more informative in detecting a two-branch solution than just 5 stars. This is evident by analyzing the distance between the model's thresholds and the first histogram's values (represented as purple vertical lines): higher NLL values are assigned to samples in which an anomaly (two-branch solution) is most likely present.\\

\begin{figure}[!htb]
    \centering
    \includegraphics[width=0.95\linewidth]{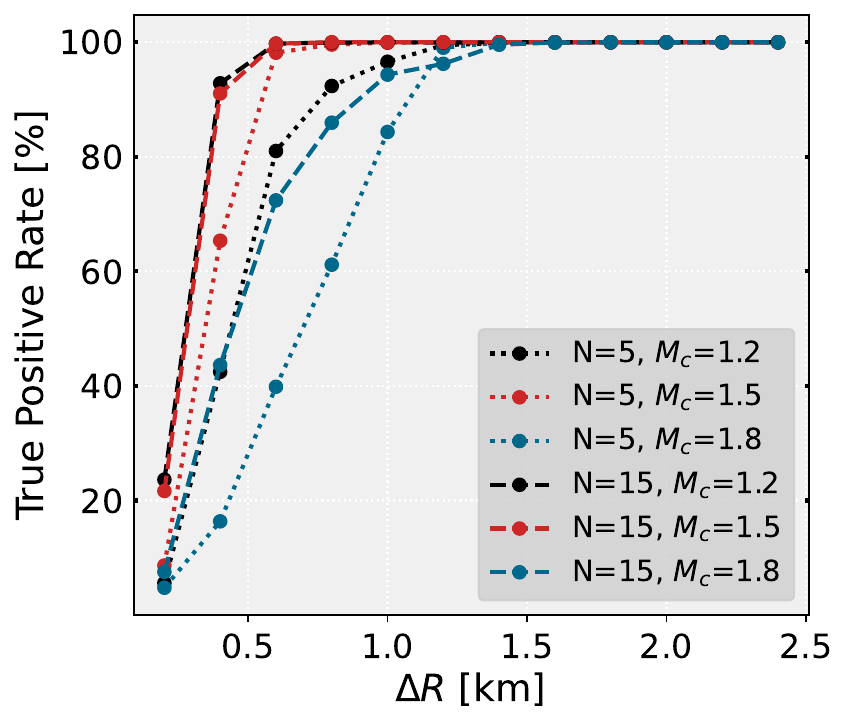}
    \caption{True positive rate (TPR), phase transition detection rate, as a function of $\Delta R$ for different $M_c/M_{\odot}$ values: $1.2$ (black), $1.5$ (red), and $1.8$ (blue) for Set 1. The use of $N=5$ or $15$ stars is represented by dashed and dotted lines, respectively.}
    \label{fig:TPR_set1}
\end{figure}

\begin{figure}[!htb]
    \centering
    \includegraphics[width=0.95\linewidth]{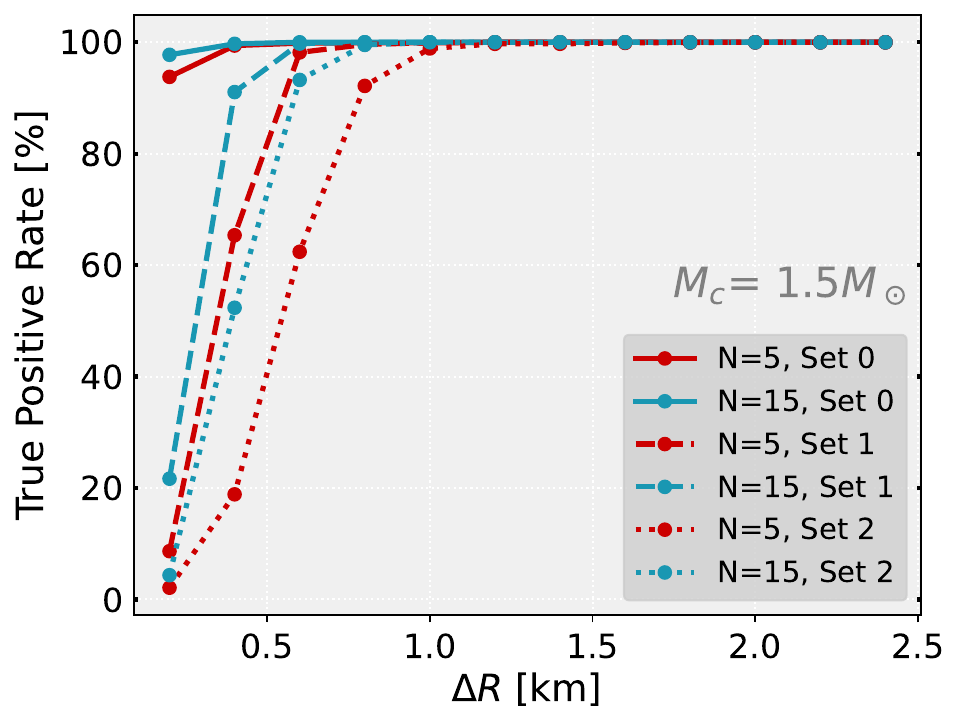}\\
    \includegraphics[width=0.95\linewidth]{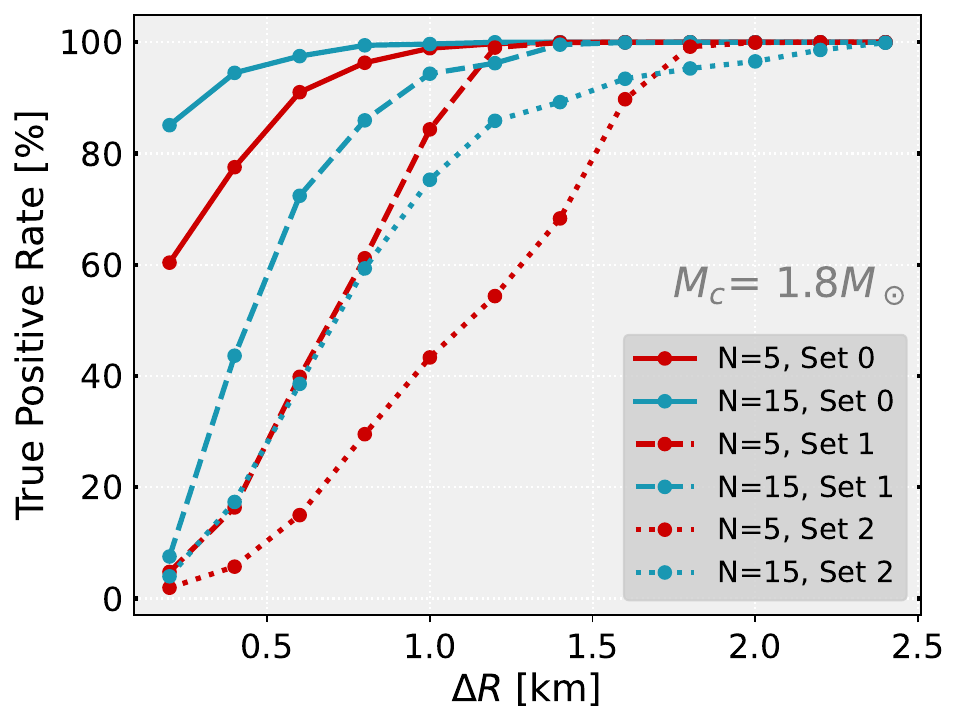}
    \caption{True positive rate (TPR) as a function of $\Delta R$ for $M_c/M_{\odot}=1.5$ (upper plot) and $1.8$ (bottom plot). The different data sets, which represent different $(M,R)$ observation uncertainties (see Table \ref{tab:sets}), are distinguished by line-type: Set 0 (solid), Set 1 (dashed), and Set 2 (dotted). 
    The use of $N=5$ or $15$ stars is represented by solid and dashed lines, respectively.}
    \label{fig:TPR_sets}
\end{figure}

For each threshold within one of the three possible critical masses, we determined how many true positives and false negatives the model detects by counting the number of values above or below the threshold across all different values of $\Delta R$. This allows us to determine the True Positive Rate (TPR)\footnote{ The TPR is determined as TPR=$TP/{(TP+FN)}$ where TP is the True positives (instances correctly predicted as positives) and FN is the False Negatives (instances incorrectly predicted as negatives). }, also known as recall -- a metric that quantifies the ability of the model to correctly identify positive instances --  for each $\Delta R$. The TPR for Set 1 is illustrated in Fig.~\ref{fig:TPR_set1} for the three critical masses and in Fig.~\ref{fig:TPR_sets} for two fixed critical masses but across all sets.

The observed increasing trend in Fig.~\ref{fig:TPR_set1}, which presents results exclusively for Set 1,  aligns with expectations -- it is easier to distinguish a two branch solution as $\Delta R$ increases.  
As expected, a larger size for the target vector ($N$) is more informative and yields larger TPR values.  Comparing the different critical masses, the model shows inferior results for $1.8 M_{\odot}$ because the phase transition occurs at a higher mass value, leading the model to statistically encounter a fewer number of hybrid stars than hadronic stars, increasing the difficulty in detecting any potential phase transition. In the $M_c=1.2 M_{\odot}$ and $1.5 M_{\odot}$ cases, we observe distinct behaviors for the two different values of $N$. For $N=5$, the TPR is higher for $1.5 M_{\odot}$ as the critical mass is positioned in the middle of the curve, making it more susceptible to variations compared to $1.2 M_{\odot}$, which tends to have only one star below the critical mass for the majority of instances.
However, the behavior is different for $N=15$. This can be attributed to the availability of more stars and the model is exposed to a greater number of instances below $1.2 M_{\odot}$. The model's increased sensitivity to detect a shifted curve, owing to its training with more information, leads to similar results for the two critical masses.\\

Figure \ref{fig:TPR_sets} shows consistently inferior results for TPR at $1.8 M_{\odot}$, in agreement with earlier findings. Furthermore, TPR values are consistently higher for $N=15$. Examining the three distinct sets, a clear pattern emerges: Set 0 exhibits higher TPR than Set 1, which, in turn, outperforms Set 2. This hierarchy is expected, as introducing more noise to the model's input makes it progressively challenging for the model to effectively detect phase transitions. The results indicate that, given 15 NS observations and $M_c=1.5 M_{\odot}$, the trained NF models are capable of detecting a phase transition with 100\% accuracy if $\Delta R > 0.4$ km (Set 0),  $\Delta R > 0.6$ km (Set 1), and $\Delta R > 0.8$ km (Set 2). The accuracy gets considerably lower for $M_c=1.8 M_{\odot}$, requiring a much larger $\Delta R$ to reach 100\% accuracy: $\Delta R > 1 $ km (Set 0),  $\Delta R >1.6$ km (Set 1), and $\Delta R > 2.4$ km (Set 2).

\begin{figure}[!htb]
    \centering
    \includegraphics[width=0.95\linewidth]{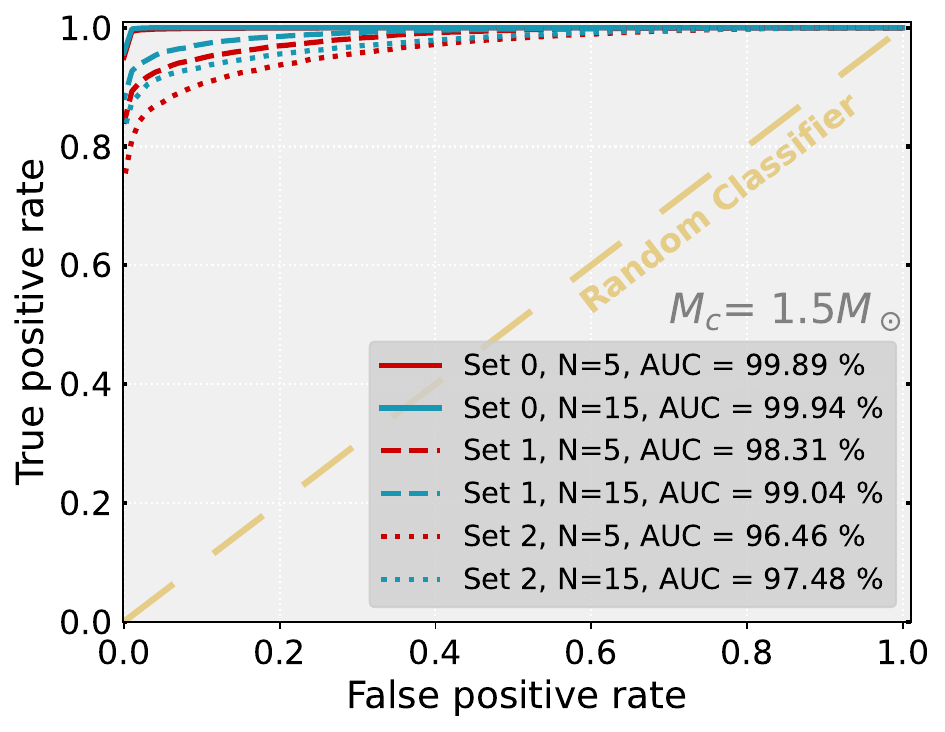}\\
    \includegraphics[width=0.95\linewidth]{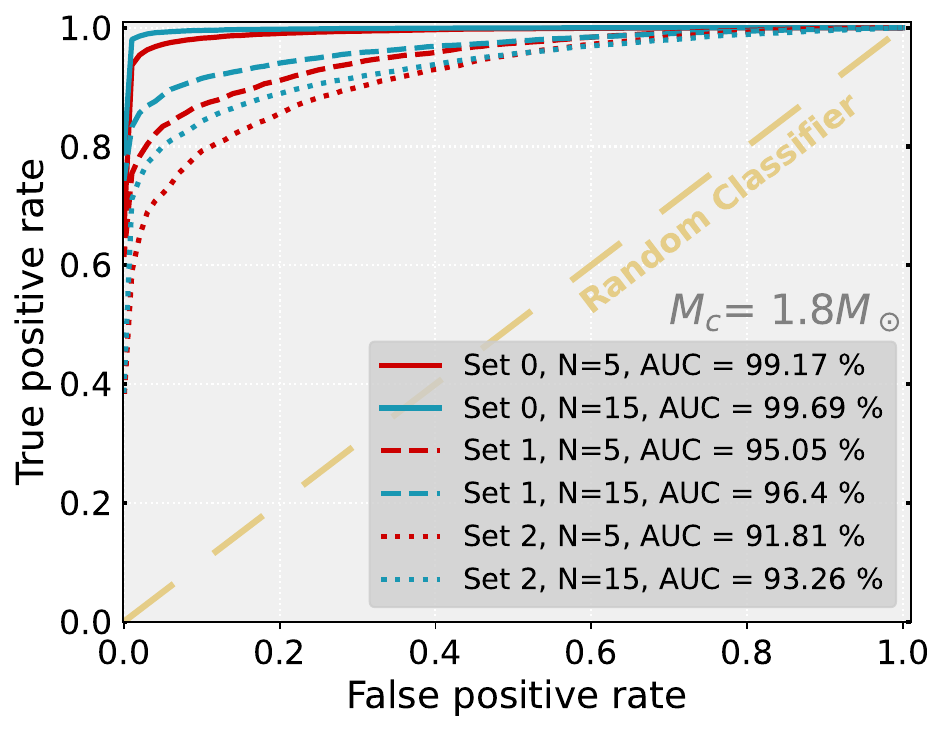}
    \caption{ROC curves illustrating the performance of the six different trained models, along with the corresponding AUC values, for the critical masses of $M_c=1.5M_{\odot}$ (top plot) and $M_c=1.8M_{\odot}$ (bottom plot). }
    \label{fig:ROC_}
\end{figure}

To evaluate our NF models using an alternative metric, we employed the Receiver Operating Characteristic (ROC) curve \cite{majnik2013roc}. Defined in the (TPR, FPR) space, where both $(0,0)$ and $(1,1)$ represent extreme classifier models, this curve provides valuable insights into the model's performance across different threshold settings. The point $(0,0)$ is achieved by increasing the threshold to classify everything as negative for the anomaly presence, resulting in both TPR and FPR being zero. The opposite occurs at $(1,1)$, where the threshold classifies everything as an anomaly, leading to both TPR and FPR being 1. The ideal point is (0,1), signifying a TPR of 1 and an FPR of 0. The ROC curve is then calculated as (TPR, FPR) for a continuous change of the threshold. An important aspect of this metric is that it provides an overview of the model's quality without the need to select a specific threshold.
The area under the curve (AUC) \cite{BRADLEY19971145} provided by ROC curve offers a straightforward classification of our models, ranging from 0 to 1 (or as a percentage). An AUC value of 1 indicates a perfect classifier, achieving a TPR of 1 with no false positives, visually represented by an ROC curve reaching the top-left corner. In contrast, a diagonal line corresponds to an AUC of 0.5, indicating a random classifier. By examining how the ROC curves approximate towards the top-left corner, we can estimate the AUC value.
We use ROC curves to classify the differences between various scenarios constructed within our different sets. In Fig.~\ref{fig:ROC_}, we present the ROC curves and the corresponding AUC values across the six distinct trained models for $M_c=1.5 M_{\odot}$ (top plot) and $M_c=1.8 M_{\odot}$ (bottom plot).
The AUC values remain larger for $M_c=1.5 M_{\odot}$, indicating that the model adeptly discerns the presence of a phase transition or anomaly with heightened precision. This pattern persists across all sets, with AUC values ranking in the order Set 0 $>$ Set 1 $>$ Set 2, further emphasising that higher noise levels in the input hinder the model's ability to detect phase transitions, and larger input sizes ($N=15$) consistently yield better performance.

\begin{figure}[!htb]
    \centering
    \includegraphics[width=0.95\linewidth]{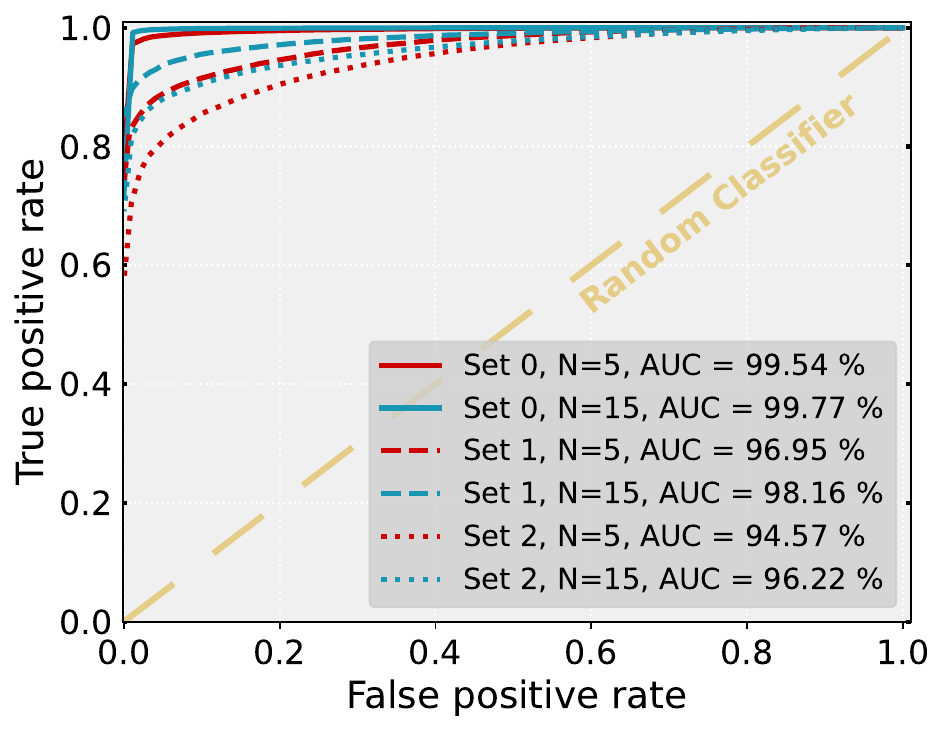}
    \caption{ROC curves illustrating the performance of the six different trained models, along with the corresponding AUC values, considering all the hypotheses for the critical masses.}
    \label{fig:ROC_t}
\end{figure}

Figure \ref{fig:ROC_t} averages each of the six distinct trained models over the $M_c$ values. Therefore, the ROC curves and the corresponding AUCs can be interpreted as an overall measure of the model's performance. The same pattern of model performance is reinforced: increasing the set size increases the performance and noisier $M(R)$ samples decrease the anomaly detection rate. This robust performance provides strong evidence supporting the model's capability to effectively detect phase transitions.\\

\begin{figure}[!htb]
    \centering
    \includegraphics[width=0.9\linewidth]{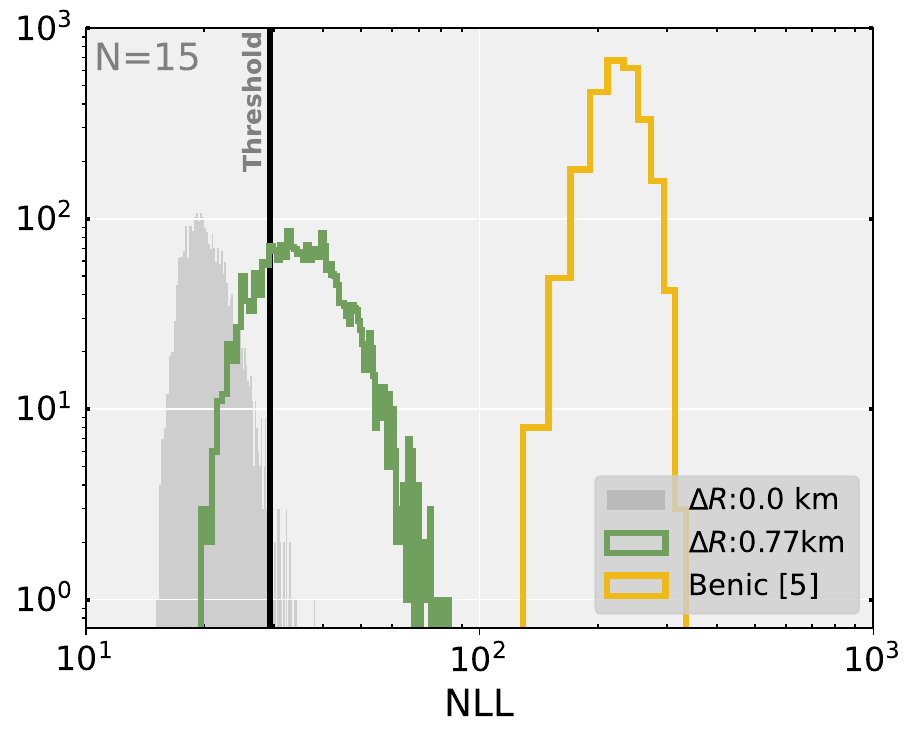}
    \caption{NLL histogram comparing a test set generated for Set 1 and $N=15$ with $(M_c,\Delta R)=(1.91 M_{\odot},0.77\text{ km})$ in green and the two-branch $M(R)$ curves derived from microscopic models \cite{Benic:2014jia} in yellow. The grey region represents the NLL of the NF model on the test set with $\Delta R=0$ (purely hadronic solution) and the vertical black line corresponds to the respective threshold value. }
    \label{fig:hist_new}
\end{figure}

Lastly, we tested the NF model trained with Set 1 and $N=15$ to a two-branch $M(R)$ curves obtained from microscopic models \cite{Benic:2014jia}, based on DD2 nuclear model \cite{Typel1999} and the Nambu--Jona-Lasinio quark model \cite{Hatsuda:1994pi}. This evaluation involved analyzing the behavior of the histogram of the NLL in relation to the predefined threshold.
To conduct this test, we adapted the microscopic model set, generating the same number of samples as in our test set with the specifications of Set 1 and $N=15$. Additionally, for a more comprehensive evaluation, we created a specific test set with $(M_c, \Delta R) = (1.91 M_{\odot}, 0.77\text{ km})$, matching the critical mass and $\Delta R$ shift of the microscopic model \cite{Benic:2014jia}.
The results in Fig.~\ref{fig:hist_new} show that the NF model detects the two-branch solution for the microscopic model (yellow) with higher confidence than our test set (green), producing NLL values that are noticeably farther from the defined threshold. 

Our two-branch solutions can be seen as a lower bound on the model performance, and any other two-branch solution construction becomes easily detectable. This is the case because we are attributing the same properties at the two branches and it is hard for the NF model to detect a phase transition from a noise sample for small $\Delta R$, which may be easily and wrongly considered from a single one-branch hadronic solution.  In the case of \cite{Benic:2014jia} results, the considerable scattering of radii values for hybrid stars and their small mass range make them easier to detect.  

\section{\label{sec:conclusions} Conclusions}

We employed NF to identify the presence of a second and disconnected branch of hybrid stars in the NS mass-radius diagram. NF are generative machine learning models capable of generating new samples and estimating the probability density of new data points. Our approach involved training NF models initially on continuous and unique hadronic branch solutions. The input samples incorporate three levels of noise (see Table \ref{tab:sets}) with two possible sample sizes (5 stars and 15 stars), resulting in the training of a total of six NF models focused on hadronic solutions.
After the training phase, we evaluated each model's capability to detect possible phase transitions, simulating diverse two-branch solutions by parameterizing them using the star mass at the transition and the radius gap, respectively $(M_c,\Delta R)$. There is no quark model underlying the hybrid branch because as our focus was solely on the general $M(R)$ structure.
The onset of the hybrid branch is marked by the critical mass value $M_c$, with the radius gap between the heaviest hadronic star and the lightest hybrid star represented by $\Delta R$. To quantify the model's ability to detect phase transitions, we approached it as a binary classification problem, defining a threshold condition based on 1\% FPR for the NLL space.\\

The key findings of our investigation highlight the model's enhanced phase transition detection accuracy. These outcomes include the model's adept response to the increase in $\Delta R$ across all scenarios, consistently superior precision with an input size of 15 stars as opposed to 5 stars, and the varied outcomes arising from distinct critical masses—highlighting a lower susceptibility to anomaly detection for $M_c\gtrsim1.8 M_{\odot}$. Furthermore, another anticipated observation was the reduction in model precision with the increase of the input noise. Furthermore, we evaluated the reliance of our NF models using two-branch $M(R)$ curves obtained from microscopic models \cite{Benic:2014jia}. As anticipated, our trained NF models demonstrated enhanced detection capabilities for these two-branch solutions when compared to our test sets. This observation aligns with the straightforward phase transition method we implemented in our study. In summary, our model demonstrated highly expected outcomes, aligning with the anticipated behavior we were expecting for each evaluation. The consistently high AUC values further underline the reliability and robustness of our approach.\\

In this study, we have not applied the NF models to real observations due to the considerable uncertainty of the present observational data, which could potentially yield inconclusive results, as shown in \cite{PhysRevC.106.065802}. 
A natural next step for future work would be to expand the data input by gravitational-wave observations, by including the measurements of tidal deformabilities (the mass-weighted tidal deformability of a binary system $\tilde{\Lambda}$, and individual components' tidal deformabilities), in order to systematically investigate regions of phase-transition EoS parameter space which may be degenerate in this type of data \cite{Soma:2023rmq,Raithel:2022efm}, and directly infer microscopic parameters, e.g. the size of the density discontinuity from $\Delta{R}$.

 A noteworthy aspect not addressed in this study is the quantification of uncertainty, a subject extensively explored in our prior two works \cite{Carvalho:2023ele,Carvalho:2024kgf}. This aspect gains increased importance for these models, as they involve multiple neural network architectures within a single forward pass, accentuating the importance of uncertainty quantification. 
 While acknowledging that implementing Bayesian Neural Networks within the framework of NF would undoubtedly pose computational challenges, it holds potential for future investigations, as exemplified in the work by \cite{hortua2020constraining}.

\begin{acknowledgments}
This work was supported by FCT - Fundação para a Ciência e Tecnologia, I.P. through the projects UIDB/04564/2020 and UIDP/04564/2020, with DOI identifiers 10.54499/UIDB/04564/2020 and 10.54499/UIDP/04564/2020, respectively, and the project
2022.06460.PTDC with the associated DOI identifier
10.54499/2022.06460.PTDC.
\end{acknowledgments}
\bibliographystyle{apsrev4-1}
%

\end{document}